\documentclass[seceq]{ptptex}

\usepackage{graphicx}

\title{
Explicit treatment of tensor force with \\
a method of Antisymmetrized Molecular Dynamics}

\author{
Akinobu \textsc{Dote}$^1$, 
Yoshiko \textsc{Kanada-En'yo}$^2$, 
Hisashi \textsc{Horiuchi}$^3$, \\ 
Yoshinori \textsc{Akaishi}$^4$ and 
Kiyomi \textsc{Ikeda}$^5$
}

\inst{
$^1$High Energy Accelerator Research Organization (KEK), Ibaraki 305-0801, 
Japan \\
$^2$Yukawa Institute for Theoretical Physics, Kyoto 606-8502, Japan \\
$^3$Department of Physics, Kyoto University, Kyoto 606-8502, Japan \\
$^4$College of Science and Technology, Nihon University, Funabashi 
274-8501, Japan \\
$^5$The Institute of Physical and Chemical Research (RIKEN), Saitama 351-0198, 
Japan
}

\abst{
In order to treat tensor force explicitly, 
we propose a microscopic model for nuclear structure 
based on antisymmetrized molecular 
dynamics (AMD). As a result of the present study, it is found that 
some extentions of the AMD method are effective to incorporate the tensor
correlation into wave functions.
Calculating the deuteron, triton and $^4$He with the extended version of AMD, 
we obtained solutions that have large contribution of the tensor force. 
By analyzing the wave funtion of $^4$He, it is found that 
its single-particle orbits are composed of 
normal-size $0s$ orbits and shrunk $p$ orbits 
so as to gain the tensor force. 
}

\begin{document}

\maketitle

\section{Introduction}

Tensor force plays important roles concerning various properties of nuclei. 
Recent studies with $ab$ $initio$ calculations show that 
the contribution of the tensor force to 
binding energies is large in $A \simeq 10$ light nuclei\cite{GFMC_2000},
as well as in few-nucleon systems.\cite{FewBody} 
Besides the binding energies, 
effects of the tensor force on single-particle energies and 
nuclear structure have been suggested for a long time 
\cite{Terasawa_LS,Akaishi-Nagata}. Recently, several studies 
showed its influences on $LS$ splitting \cite{Myo_tensor},  
$GT$ transitions and  magnetic moments \cite{Sugimoto:RMF}. 
These facts imply that direct effects of the tensor force may be revealed 
in some properties of nuclear structure. 

The tensor force causes a characteristic correlation between two nucleons, 
which is called the {\it tensor correlation}.  
In the deuteron, the tensor force gives a large attraction due to the coupling 
of $^3S_1$ with $^3D_1$. Also, in $^4$He,  
the [$JLS=022$] component is mixed to the dominant $(0s)^4$ component due to the 
tensor force. 
Since the tensor correlation between two nucleons is varied,  
depending on the environment surrounding them, the contribution of the tensor 
force is sensitive to the nuclear density and structure. 
According to analyses with the $g$ matrix,  
as the density increases in nuclear matter, 
the contribution of the tensor force is suppressed ({\it tensor suppression}).  
This is a major cause of the nuclear saturation property.\cite{Bethe} 
Reversely, as the density decreases, 
the tensor suppression becomes small. 
Therefore, the tensor force is considered to be 
important in light nuclei because they have a relatively large surface region, 
being low density. 
In Ref. [\ref{Akaishi-Nagata}], it is emphasized that 
the renormalization of the tensor force 
to an effective $^3E$ central force is very sensitive 
to the starting energy as well as the density. 
Because the starting energy is strongly dependent on the 
structure of the total system, the tensor force gives a structure dependence 
to the effective interaction. Thus, 
the tensor force might be closely related to a variety of structures 
realized in light nuclei such as a clustering structure.
Moreover, it is natural to expect
that the tensor force may provide clarity to 
the problems of unstable nuclei, which 
usual approaches with effective nuclear forces sometimes fail to describe.

There are many kinds of model calculations that are useful 
for systematic studies of various nuclei in wide mass-number regions. 
They, however, have avoided any explicit  
treatment of the tensor force because of its difficulty. 
In such model calculations, one uses effective interactions 
that are either derived from realistic forces theoretically 
by renormalizing the tensor force into central and $LS$ forces, 
or phenomenologically determined. 
In usual cases, 
the density dependence arising from the tensor force is simulated 
by phenomenological density-dependent or three-body forces.  
Such model calculations with effective interactions have succeeded to 
reproduce 
and explain many physical quantities of various nuclei. 
However, there remain some phenomena that those model calculations
failed to reproduce without using system-dependent effective forces.
The parity inversion problem of $^{11}$Be and the binding mechanism 
of $^{11}$Li are typical examples. Therefore, it is natural to expect that
the effect of the tensor force might be a key to solve these 
problems.\cite{Myo_11Li} 
Moreover, another drawback of the usual model calculations is that
one can not discuss the effect of the tensor force on physical 
quantities in an explicit way, because the obtained wave functions do not 
contain a tensor correlation.  

Our aim is to treat the tensor force as directly as possible 
in the model calculation. 
We would like to systematically investigate its contribution 
to the binding of various nuclei and the effects of the tensor correlation 
on the nuclear properties. 
We mention here a repulsive core of the nuclear force, which  
is another characteristic of a realistic nuclear force,  
and is not explicitly contained in the usual effective nuclear force 
as well as the tensor force. 
Recent developments in treating the repulsive core and the tensor force 
in a model calculation can be seen in 
the unitary correlator method\cite{Neff_shell}, 
the $g$-matrix method\cite{AkaishiTensor} and the $V_{low-k}$ method\cite{Vlowk}.
Since the repulsive core should be less 
state-dependent than the tensor force, we consider that it can be smoothed out 
for the present purpose.
We adopt an effective force where the repulsive core in the central force is
smoothed by the $g$-matrix method in the present study.
 
To accomplish our purpose, we need a wave function 
that can contain the tensor correlation. 
We start from a mean-field approach, considering its applicability to 
wide mass-number regions. 
As already known, ordinary mean-field wave functions are 
insufficient to treat the tensor force. 
From the operator form of the tensor force, one can easily expect that  
two extensions are needed. 
The first one is 
i) {\it a parity-violating mean field and a strong correlation between spin 
and space.} 
The other point is 
ii) {\it flexibility of the isospin wave function.} 
The tensor force contains a parity-odd operator as a dominant term  
in a mean-field picture, 
and it also includes a charge-exchange term. 
Therefore,  
it is difficult to treat the tensor force within the usual mean-field approaches.
In fact, the expectation value of the tensor operator vanishes in a model space 
of a simple mean-field.
The importance of the parity-violating mean field and charge mixing 
are already pointed out in the Ref. [\ref{Sugimoto:CPPHF}]. 

Several groups are currently challenging this theme. 
Sugimoto {\it et al.} are investigating the contribution of the tensor force 
in various nuclei, based on 
the relativistic\cite{Sugimoto:RMF} and non-relativistic\cite{Sugimoto:CPPHF} 
mean-field approaches.  
Myo {\it et al.} are investigating the effect of a tensor correlation 
in light unstable nuclei with an extended shell model.\cite{Myo_tensor} 
We, here, adopt an antisymmetrized molecular dynamics (AMD) method\cite{AMD:Enyo} 
as a starting point of our model. 
Due to the simplicity of its framework, AMD can be easily improved,  
while corresponding to our requirements.\cite{AMD:Enyo-VAP, AMD-HF:Dote2, 
AMD:Kimura, AMD:Taniguchi} 
Since its wave function is simply composed of Gaussian wave packets, 
we can easily perform various projections, such as parity, angular momentum 
and charge number projections. 
The parity-violating mean field has already been realized 
by the parity projection in AMD. 
Other advantages of the AMD are that it does not rely on any model assumptions 
on nuclear structure, such as the existence of clusters 
and nuclear deformation, and that it is applicable to general nuclei over 
a wide mass-number region. 
It has succeeded to describe various structures and cluster aspects
of light and middle nuclei.\cite{AMD:Enyo, AMD:Enyo-VAP, AMD-HF:Dote2, 
AMD:Kimura, AMD:Taniguchi}
We, therefore, expect that a model based on AMD 
can directly reveal a relation 
between the tensor force and various structures of light nuclei, 
including clustering structure, as investigated in Ref. [\ref{Akaishi-Nagata}]. 
On the other hand, in Refs.\cite{Myo_tensor,Sugimoto:RMF,Sugimoto:CPPHF}, the 
model wave functions are based on mean-field approaches, and the 
single-particle orbits in $^{4}$He are assumed to be linear 
combination of $s$ and $p$-orbits. 
Based on the AMD method, we will see how the tensor correlation is 
realized in the wave functions without relying on such a model assumption.

In the present paper, we employ effective interactions 
that involve the tensor term explicitly, and extend 
the framework of AMD so that it can gain the contribution of 
the tensor force. 
We adopt an effective force that is derived from 
the Tamagaki potential\cite{Tamagaki}  
by smoothing the repulsive core of the central term,
and also test another effective force: the Furutani-Tamagaki 
potential\cite{Furutani}.
We calculate the deuteron, triton and $^4$He 
with an extended version of AMD. We show the effects of 
i) {\it a parity-violating mean field and a strong correlation between spin 
and space} and 
ii) {\it the flexibility of the isospin wave function} 
in practical calculations. Moreover, we propose 
additional extensions of the AMD to gain the tensor force more effectively;  
 iii) {\it linear combination of the wave packets that have different 
width parameters} and iv) {\it charge-, parity- and 
angular-momentum-projections}.  
Finally, we analyze the properties of the wave functions of the deuteron and $^4$He 
obtained by the extended AMD, while focusing on the tensor correlation.

This paper is organized as follows.  
In section 2, we explain the extensions of the AMD framework. 
We investigate the effect of these extensions in the deuteron, triton and $^4$He 
in section 3. Analysis of the obtained wave function of $^4$He 
from the viewpoint of single particle levels is shown in section 4. 
In section 5, we give a summary and some discussions 
on further developments.

\section{Formalism}

\subsection{Hamiltonian and tensor force}
 
The Hamiltonian used in this paper is 
\begin{eqnarray}
H = \sum_{i=1}^A t_i + \sum_{i<j=1}^A \left( v^C_{ij} + v^T_{ij} + 
v^{Coulomb}_{ij} \right) - T_{CM}.   
\end{eqnarray}
Here, $t_i$ is a kinetic energy term,  
$\frac{{\boldmath{\mbox{$p$}}}_i^2}{2m_N}$  
(${\boldmath{\mbox{$p$}}}_i$, momentum operator; $m_N$, nucleon mass);  
$v^C_{ij}$, $v^T_{ij}$ and $v^{Coulomb}_{ij}$ indicate 
the central force, tensor force and Coulomb force, respectively. 
The center-of-mass motion energy, $T_{CM}=\frac{1}{2Am_N} 
\left( \sum_{i=1}^A \boldmath{\mbox{$p$}}_i \right)^2$, is removed. 
We employ two kinds of effective interactions, the Furutani-Tamagaki potential 
(F-T potential)\cite{Furutani} 
and  the Akaishi(T) potential (AK(T) potential), as a set of $v^C_{ij}$ and 
$v^T_{ij}$. 
Both potentials are described by a superposition of 
several range Gaussian wave packets: 
\begin{eqnarray}
v^C_{12} & = & \sum_{n=1}^{N} 
\left\{ C^{1E}_n P(^1E) + C^{3E}_n P(^3E) + C^{1O}_n P(^1O) + C^{3O}_n P(^3O)   
\right\} \exp \left[ -(r/b_n)^2 \right], \label{Vc} \\
v^T_{12} & = & \sum_{n=1}^{N}  \; T^{3E}_n P(^3E) \; r^2 S_{12} \; 
\exp \left[ -(r/b_n)^2 \right],  
\label{Vt}
\end{eqnarray} 
where $r=|\mbox{\boldmath $r$}_1-\mbox{\boldmath $r$}_2|$. 
$N=3$ in the F-T potential and $N=10$ in the AK(T) potential. 
$P(X)$, $X= \,^1E, \,^3E, \,^1O, \,^3O$, is a projection operator and 
$S_{12}$ is a tensor operator, 
$S_{12}= 3 (\mbox{\boldmath $\sigma$}_1 \cdot \mbox{\boldmath $r$}) 
(\mbox{\boldmath $\sigma$}_2 \cdot \mbox{\boldmath $r$}) / r^2 - 
(\mbox{\boldmath $\sigma$}_1 \cdot \mbox{\boldmath $\sigma$}_2)$, 
where $\mbox{\boldmath $r$}=\mbox{\boldmath $r$}_1-\mbox{\boldmath $r$}_2$. 
The radial part of the 
tensor force in both potentials is represented by a linear combination of 
$r^2 \times$ Gaussians. 
The interaction parameters, $\{ C^X_n, T^{3E}_n, b_n\}$, in the F-T potential are 
given in Ref. [\ref{Furutani}], and 
those in the AK(T) potential are given in Appendix. 
The Coulomb force, $v^{Coulomb}_{ij}$, is expressed by the superposition of 
seven range Gaussian wave packets. 

\begin{figure}[t]
\caption{(a) Furutani-Tamagaki potential and (b) Akaishi(T) potential. 
The $^1E$ and $^3E$ central potential and the $^3E$ tensor potential 
are shown as 'Central 1E', 'Central 3E' and 'Tensor 3E' in each panel, 
respectively. Both potentials are multiplied by $r^2$. \label{Potentials}}
\begin{center}
\begin{minipage}[t]{6.5cm}
\begin{center}
\includegraphics[width=6.5cm]{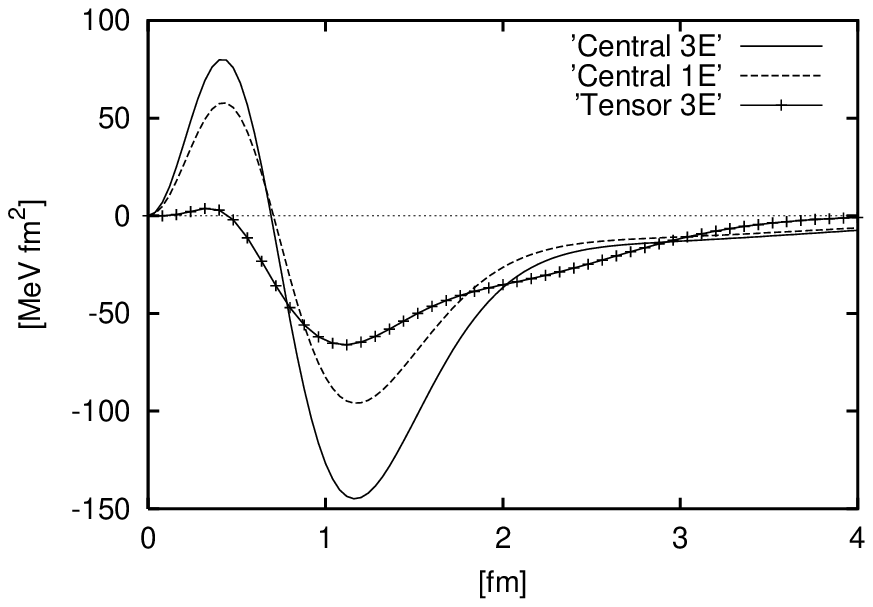}\\
(a) Furutani-Tamagaki potential 
\end{center}
\end{minipage}
\hspace{0.5cm}
\begin{minipage}[t]{6.5cm}
\begin{center}
\includegraphics[width=6.5cm]{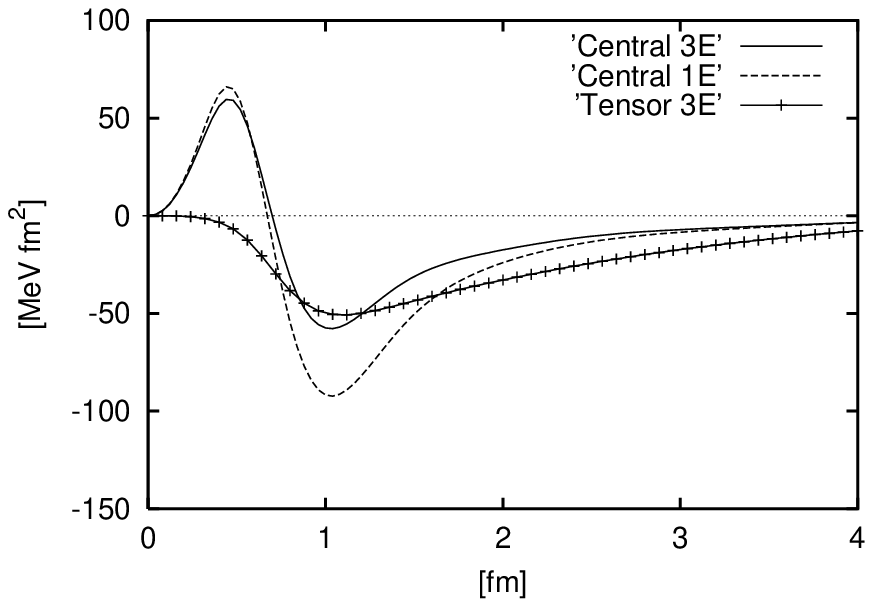}\\
(b) Akaishi(T) potential
\end{center}
\end{minipage}
\end{center}
\end{figure}

The F-T potential was used in a  cluster model study of $^3$He+$p$ 
scattering, and succeeded in reproducing scattering data, including 
polarization quantities and the spectra of $^4$Li.\cite{Furutani} 
It can also reproduce 
the binding energy of $^4$He with a harmonic-oscillator $(0s)^4$ wave 
function. 
The AK(T) potential is derived from the Tamagaki potential (OPEG),\cite{Tamagaki} 
which is a realistic nucleon-nucleon potential. 
The central part of the AK(T) potential is obtained by 
smoothing the repulsive core of the central part of OPEG with the method of 
$g$-matrix theory, 
whereas the tensor part remains to be the same as the original one of OPEG. 
In other words, the tensor force is not renormalized 
into the central one when constructing the AK(T) potential. 
The details of its derivation are explained in 
Ref. [\ref{AkaishiTensor}].\footnote{
In Ref. [\ref{AkaishiTensor}], the AV8' potential\cite{AV8p} is used and 
several values of the cut-off momentum, $k_Q^{(D)}$, 
for the tensor force are investigated. The AK(T) potential corresponds to the case 
that the Tamagaki potential is used instead of the AV8' potential and 
$k_Q^{(D)}=\infty$ fm$^{-1}$.} 
The $^1E$ and $^3E$ central forces and the $^3E$ tensor force 
in both potentials are depicted in Figure \ref{Potentials}.
In the F-T potential, the $^3E$ central force is dominant, while 
the tensor force is smaller. 
On the other hand, the $^3E$ central force in the AK(T) potential 
is smaller than that in the F-T potential. 
The tensor force has almost the same magnitude as the $^3E$ central force 
in the AK(T) potential. 

\subsection{Extended framework of AMD}

The details of the usual AMD framework are given in Refs. [\ref{AMD_org}] 
and [\ref{AMD_rev}]. 
In order to incorporate the effect of the tensor force efficiently, 
we improve 
the single-nucleon wave functions, $|\varphi_i \rangle$, by superposing Gaussian 
wave packets, 
as follows: 
\begin{eqnarray}
|\varphi_i \rangle & = & \sum_{\alpha=1}^N 
\left( \frac{2\nu_\alpha}{\pi} \right)^{3/4} 
C^i_\alpha 
\exp\left[ -\nu_\alpha \left( \mbox{\boldmath $r$} - 
\frac{\mbox{\boldmath $Z$}^i_\alpha}{\sqrt{\nu_\alpha}}\right)^2 \right] |
\beta^i_\alpha \rangle 
|\gamma^i_\alpha \rangle,  \label{Eq:SNwfn}\\
& & |\beta^i_\alpha \rangle= \left( \frac{1}{2} + \beta^i_\alpha \right) |
\uparrow \rangle + 
\left( \frac{1}{2} - \beta^i_\alpha \right) |\downarrow \rangle, \\
& & |\gamma^i_\alpha \rangle= \left( \frac{1}{2} + \gamma^i_\alpha \right) |
p \rangle + 
\left( \frac{1}{2} - \gamma^i_\alpha \right) |n \rangle,  \label{eq:iso}
\end{eqnarray}
where $|\beta^i_\alpha \rangle$ is an intrinsic-spin wave function and 
$|\gamma^i_\alpha \rangle$ is an isospin wave function. 
Coefficients $C^i_\alpha$, $\beta^i_\alpha$, $\gamma^i_\alpha$ and the Gaussian 
center 
$\mbox{\boldmath $Z$}^i_\alpha$ are complex variational parameters. 
This single-nucleon wave function, $|\varphi_i \rangle$, is more sophisticated 
than that in our previous study\cite{AMD-HF:Dote2}, 
where only the spatial part of a single nucleon wave function was expressed by 
the superposition of Gaussian wave packets, regarding several points. 
In the present wave function, not only the centers 
$\{ \mbox{\boldmath $Z$}^i_\alpha \}$, 
but also the width parameters $\{ \nu_\alpha \}$ of the wave packets 
are different from each other. 
Moreover, the variational intrinsic-spin wave function, 
$| \beta^i_\alpha \rangle$, 
is independently given for each wave packet. Since the present single-nucleon 
wave function, $|\varphi_i \rangle$, is possible 
to describe  the strong correlation between the spatial part and the 
spin part, 
it can effectively take the contribution from 
the term $ \mbox{\boldmath $\sigma$} \cdot \mbox{\boldmath $r$}$ 
in the tensor force, as mentioned in the previous section. 
In the isospin part, we also use the variational isospin wave function, 
$|\gamma^i_\alpha \rangle$, as shown in Eq. (\ref{eq:iso}), so that it 
can contain both the proton and neutron components, 
though it is fixed to be a proton or neutron state in the usual AMD. 
Due to the flexibility of the isospin wave function, 
the present wave function can take 
effect of the term $\mbox{\boldmath $\tau$}_1 \cdot 
\mbox{\boldmath $\tau$}_2$, 
especially $\tau_{1+} \tau_{2-}$ and $\tau_{1-} \tau_{2+}$, 
in the tensor force. 
This corresponds to point ii) mentioned in the previous section. 

In our treatment, the total wave function is constructed 
from an intrinsic wave function, $|\Phi \rangle$, with 
various projections, where $|\Phi \rangle$ is a Slater determinant 
of $\{|\varphi_i\rangle\}$: 
$|\Phi \rangle = \det [|\varphi_i\rangle]$. 
We performed parity projection ($P_{\cal P}$) and angular momentum 
projection ($P_J$) 
in the same way as for the usual AMD calculation for a nuclear structure study. 
In addition, 
we need {\it charge-projection} ($P_{T_z}$) in the present framework, 
since $|\Phi \rangle$ is a charge-mixed state, as shown in 
Eq. (\ref{eq:iso}). 
The charge projection is done by rotating around the z-axis in isospin space, 
\begin{eqnarray}
|P_{T_z} \, \Phi \rangle = \int d\alpha \exp 
\left[ i \alpha \left( \hat{T}_z -M \right) \right] \; 
|\Phi \rangle .
\end{eqnarray} 
Thus, the total wave function in the present study is the 
parity-, angular-momentum- and charge-number-eigen wave function, 
\begin{eqnarray}
|P_J P_{T_z} P_{\cal P} \, \Phi \rangle.
\end{eqnarray}

Our procedure of energy variation and  projections is as follows. 
We employ $|P_{T_z} P_{\cal P} \, \Phi \rangle$  as a trial wave function 
in the energy variation. 
The complex variational parameters, 
$\{ X^i_\alpha \}=\{ C^i_\alpha, \, \mbox{\boldmath $Z$}^i_\alpha, \,  
\beta^i_\alpha. \,  
\gamma^i_\alpha \}$, are determined so as to minimize the energy, 
\begin{equation}
{\cal H} = \frac{\langle P_{T_z} P_{\cal P} \, \Phi |H|P_{T_z} P_{\cal P} \, 
\Phi \rangle}
{\langle P_{T_z} P_{\cal P} \, \Phi |P_{T_z} P_{\cal P} \, \Phi \rangle},  
\end{equation}
by the ``frictional cooling method with constraints,'' 
which is one of  the energy-variation methods. 
After the energy-variation, $|P_{T_z} P_{\cal P} \, \Phi \rangle$ is 
projected onto the 
eigen-state of the total angular momentum; $|P_J P_{T_z} P_{\cal P} \, \Phi 
\rangle$. 
We evaluate various quantities with the wave function 
$|P_J P_{T_z} P_{\cal P} \, \Phi \rangle$. 

We here comment on the practical scheme of 
the frictional cooling method with constraints. 
As mentioned in Ref. [\ref{AMD-HF1}], we need a constraint condition 
to fix the position and momentum of the center of mass 
($\langle {\bf R}_G \rangle$ and $\langle {\bf P}_G \rangle$) to the origin, 
when a single nucleon wave function is represented by the superposition of 
several Gaussian wave packets. 
In the present study, such several constraint conditions 
are satisfied by adding harmonic-oscillator-type potentials to 
the energy, ${\cal H}$, in the frictional cooling equation: 
\begin{eqnarray}
\dot{X}^i_\alpha & = & (\lambda + i\mu) \frac{\partial {\cal H}'}{\partial 
X^{i*}_\alpha} 
\hspace{0.5cm} {\rm and \; \; C.C.}, \\
& & {\cal H}' = {\cal H} + C_1 \left\{\langle {\bf R}_G \rangle^2 + 
\langle {\bf P}_G \rangle^2 \right\} + \sum_{a=2}^M C_a (W_a - W_a^0)^2 , 
\label{constraint}
\end{eqnarray}
where $W_a$ and $W_a^0$ indicate an expectation value of some operator and 
a constant value, respectively. 
By adopting adequately large positive constant values of $C_1$ and $C_a$ 
in this equation,  
we can obtain 
an optimum solution that satisfies several constraint conditions 
$\{W_a=W_a^0\}$ 
after frictional cooling. 
This method works much better than 
the Lagrange-multiplier method.\cite{AMD-HF:Dote2, AMD-HF:Dote1}

\section{Practical results for deuteron and $^4$He}

As explained in \S 2, we drastically extend 
the AMD wave function for treating the tensor force. 
In this section, we show the effect of each extension on the deuteron and $^4$He. 

We consider four extensions: 
i) a parity-violating mean field and a strong correlation between spin and space, 
ii) flexibility on the isospin wave function, 
iii) different width wave packets 
and 
iv) charge-, parity- and angular momentum projections.
i) and ii) are attributed to the operator form of the tensor force, 
$(\mbox{\boldmath $\tau$} \cdot \mbox{\boldmath $\tau$}) \; 
(\mbox{\boldmath $\sigma$}_1 \cdot \mbox{\boldmath $r$}) 
(\mbox{\boldmath $\sigma$}_2 \cdot \mbox{\boldmath $r$})$.
Extension i) is effective to 
the operator $\mbox{\boldmath $\sigma$}_i \cdot \mbox{\boldmath $r$}$,    
because this operator changes the parity of a single-particle state and 
leads to a strong correlation between spin and space. 
In the AMD method, the former part of i) has already been realized 
because the parity projection is usually performed. 
The latter part of i) is taken into account in the present framework 
by the superposition of 
Gaussian wave packets that have different spin wave functions. 
The other operator, $\mbox{\boldmath $\tau$}_1 \cdot \mbox{\boldmath $\tau$}_2$,  
causes charge-mixing in the single-particle wave function, which is 
treated 
by extension ii). 
We first examine the influence of the tensor force on the properties of 
the wave function with these extensions. In other words,
we look into the perturbative change from the usual model calculations obtained 
by an effective central-type interaction. Therefore, 
we use the F-T potential, where the tensor term is added to a 
central interaction, 
because such properties as the binding energy and radius 
of $^4$He are reproduced with the central term in the F-T potential 
by simple AMD calculations without any extensions.
Next, we 
adopt the AK(T) potential, which is derived by smoothing out the hard core of the 
central term in the $g$-matrix theory 
based on a realistic nucleon-nucleon potential.
In practical calculations with the AK(T) potential, we show that 
 extensions iii) and iv) are important to describe the 
properties of $^4$He.

In \S\S3.1 and \S\S3.2, we investigate the effects of 
 extensions i) and ii) with the F-T potential, in a simplified model space 
where the width parameters of all wave packets 
are set to a common value, namely $\nu_\alpha=\nu$. 
The effect of extension iii) by using different width parameters 
is investigated in \S\S 3.3, with the AK(T) potential. 
In all calculations, 
the deuteron and $^4$He are projected onto $J^\pi=1^+$ and $0^+$, respectively.

\subsection{Superposition of Gaussian wave packets with different spins 
and parity projection}

\begin{figure}[t]
\caption{(a) Binding energy and expectation value of the tensor force ($V_T$) 
in the deuteron as a function of the number ($N$) of wave packets 
superposed to compose a single nucleon wave function, as described in 
Eq. (\ref{Eq:SNwfn}). 
(b) Expectation value of squared orbital angular momentum in deuteron 
for various $N$.  
The solid (dashed) lines indicate the results with independent (common) 
intrinsic spins for the wave packets in each single nucleon wave function. 
Here, the F-T potential is used for both (a) and (b). 
The details are explained in the text. \label{BE_Tensor_L2_D}}
\begin{center}
\begin{minipage}[t]{6.5cm}
\begin{center}
\includegraphics[width=6.5cm]{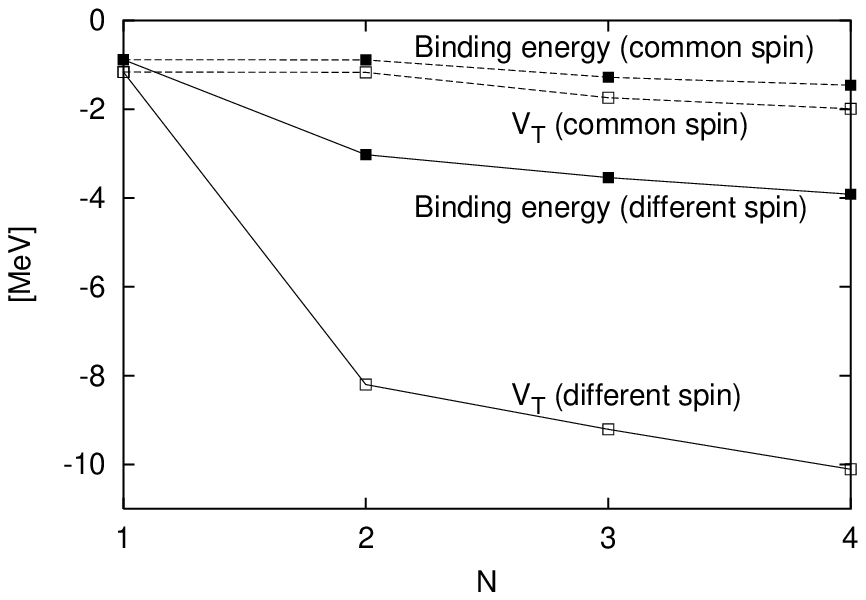}\\
(a)
\end{center}
\end{minipage}
\begin{minipage}[t]{6.5cm}
\begin{center}
\includegraphics[width=6.5cm]{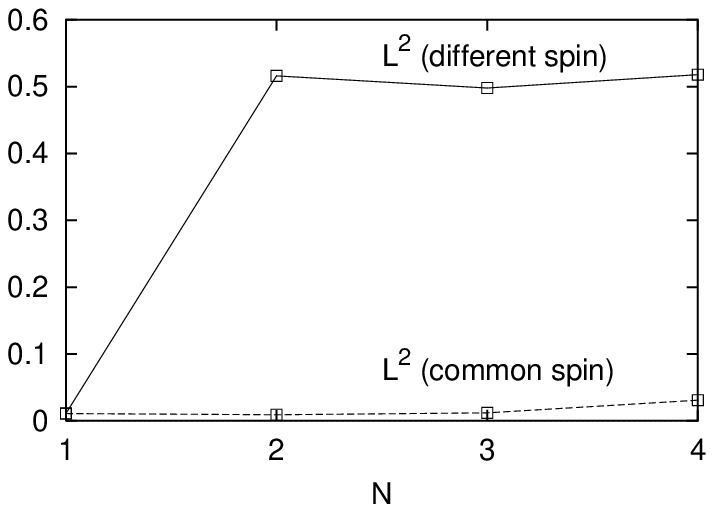}\\
(b)
\end{center}
\end{minipage}
\end{center}
\end{figure}

As mentioned before, in order to treat 
 point i) a parity-violating mean field and strong spin-space correlation 
in the AMD framework, the parity projection and the superposition of wave 
packets that have 
independent spins are needed. 
We investigated their effects by calculating the deuteron with the F-T potential. 
Fig. \ref{BE_Tensor_L2_D}(a) shows the binding energy and the expectation 
value of the 
tensor force ($V_T$) for various numbers ``$N$'' 
of wave packets superposed in $|\varphi_i\rangle$. 
Here, we have fixed each isospin wave function to be a proton or neutron; 
$|\gamma^i_\alpha \rangle = | p \rangle \; {\rm or} \; | n \rangle$ in 
Eq. (\ref{eq:iso}). 
``common spin'' (dashed line) indicates that the common spin wave function 
is used for all wave packets in each $|\varphi_i\rangle$,  
namely $|\beta^i_\alpha \rangle = |\beta^i \rangle$ in Eq. (\ref{Eq:SNwfn}), 
whereas 
``different spin'' (solid line) indicates that 
each wave packet independently has a different spin wave function, 
as shown in Eq. (\ref{Eq:SNwfn}).  
 
As shown in the figure, the expectation value of the tensor force, which 
we denote as $|V_T|$, is nearly equal to zero when the superposition 
of wave packets is not performed ($N=1$).   
In the case that different spin wave functions are used for each wave packet, 
$| V_T |$ is increased and the binding energy is gained by the superposition   
(see the $N \geq 2$ region of ``different spin'' in 
Fig. \ref{BE_Tensor_L2_D}(a)).   
A drastic energy gain of the $V_T$ term is seen when the 
number of wave packets ($N$) is changed from 1 to 2. 
On the other hand, when we use a common spin wave function for each 
wave packet, 
$| V_T |$ remains to be nearly zero, even if 
more than two wave packets are used ($N \geq 2$ and ``common spin''). 
In order to see the mixing ratio of high-angular-momentum components, we 
show the expectation value of the squared orbital angular momentum, 
$\mbox{\boldmath $L$}^2 \equiv \langle \hat{\mbox{\boldmath $L$}}^2 \rangle$, 
of the total system in Fig. \ref{BE_Tensor_L2_D}(b). 
Corresponding to the behavior of $V_T$, 
$\mbox{\boldmath $L$}^2$ is also drastically 
deviated from zero in the case that a different spin wave function is 
used (solid line). 
This deviation of $\mbox{\boldmath $L$}^2$ from zero 
indicates a mixing of the $D$-state, which is consistent with the usual 
understanding of the deuteron, that 
the tensor force is gained by $^3S_1$-$^3D_1$ coupling. 
Thus, we found that, in the description of a single nucleon wave function, 
$|\varphi_i\rangle$,   
the superposition of more than two Gaussian wave packets 
with independent spin wave functions 
is effective to gain the tensor force in the AMD framework. 

Since the operator $\mbox{\boldmath $\sigma$}_i \cdot \mbox{\boldmath $r$}$  
changes the parity of a single-particle state, 
a parity-violated single-particle state 
is necessary to incorporate the effect of this operator. 
In order to properly take into account the parity violation of 
single-particle wave functions, the parity projection before the variation 
of the total system is important. 
We show the effect of the parity projection in the deuteron in 
Table \ref{Deuteron_PP}. 
In the upper row (``w''), we show the result with the parity 
projection, which corresponds to $N=4$ wave packets with different spins 
in Fig. \ref{BE_Tensor_L2_D}(a). 
The lower row (``w/o'') shows the expectation values for the wave function 
obtained 
by switching off the parity projection. 
The contribution of the tensor force without the parity projection ($-1.5$ MeV) 
is much smaller than that of the result with the parity projection 
($-10.1$ MeV). 
$\mbox{\boldmath $L$}^2$ is also reduced from 0.52 to 0.05, 
which means a decrease of the $D$-state mixing by switching off the 
projection. 
Thus, we confirm the importance of the parity projection. 
For the deuteron calculated with parity projection,
its single nucleon wave function is actually a parity-violated state 
with the mixing of $s$- and $p$-orbits, as shown in \S 4. 

\begin{table}[b]
\caption{Effect of parity projection in the deuteron, which is calculated with 
the F-T potential. \label{Deuteron_PP}
The upper (lower) row shows the result obtained with (without) parity 
projection.
$T$, $V_C$, $V_{T}$ and B.E. are 
the kinetic energy, central force, tensor force and total binding energy, 
respectively. 
They are in units of MeV. 
$\mbox{\boldmath $L$}^2$ is the expectation value of the squared orbital 
angular momentum. }
\begin{center}
\begin{tabular}{c|ccccc}
\hline \hline
Parity projection & $T$ & $V_C$ & $V_{T}$ & B.E. & $\mbox{\boldmath $L$}^2$ \\
\hline
w    & 24.3 & $-$18.1 & $-$10.1 & 3.9 & 0.52 \\
w/o  & 16.0 & $-$15.7 & $-$1.5  & 1.2 & 0.05 \\
\hline
\end{tabular}
\end{center}
\end{table}

\subsection{Flexibility on the isospin wave function and charge projection}

We calculate the $^4$He by using F-T potential, and 
investigate the effect of the point ii) flexibility of the isospin 
wave function. The results are summarized in Table \ref{4He_cal1}. 
In calculations except for ``Usual AMD'', each single nucleon wave function is 
described by four wave packets ($N=4$) with independent intrinsic spin 
functions 
(different spins). 
The top row (``Usual AMD'') is obtained by 
the ordinary version of AMD, in which each nucleon is described with a 
single wave packet,  
and parity and angular-momentum projections are performed. 
The middle row (``Isospin fixed'') is obtained 
by a calculation in which each isospin wave function, 
$|\gamma^i_\alpha \rangle$, is fixed to be a proton or neutron, 
$|\gamma^i_\alpha \rangle = | p \rangle \; {\rm or} \; | n \rangle$,  
in the same way as the calculation of the deuteron.  
The bottom row (``Isospin free'') is obtained 
by a calculation in which $|\gamma^i_\alpha \rangle$ can describe 
a proton-neutron mixed state,  
$|\gamma^i_\alpha \rangle= \left( \frac{1}{2} + \gamma^i_\alpha \right) |
p \rangle + 
\left( \frac{1}{2} - \gamma^i_\alpha \right) |n \rangle$, with the variational 
parameter 
$\gamma^i_\alpha$, as shown in Eq. (\ref{eq:iso}).  
By comparing these results, 
$|V_T|$ and $\mbox{\boldmath $L$}^2$ are larger in the case of ``Isospin free'' 
than in the case of ``Isospin fixed''. 
Therefore, it is found that the use of the isospin-mixed single 
nucleon wave function is important to gain the tensor force in $^4$He. 

In the previous and present subsections, 
we qualitatively show that 
two extensions of the AMD framework, 
i) a parity-violating mean field and a strong correlation between spin and space 
 and  
ii) flexibility of the isospin wave function,    
are necessary to treat the tensor force. 
However, the results obtained by the F-T potential still fail to quantitatively 
reproduce 
the experimental values of such properties 
as the binding energies and the root-mean-square radii of these nuclei.  
Both of the deuteron and $^4$He are overbound and overshrunk, 
as can be seen in Table \ref{Deuteron_PP} and \ref{4He_cal1}. 

\begin{table}[b]
\caption{Results of $^4$He with the F-T potential. \label{4He_cal1}
$T$, $V_C$, $V_{LS}$, $V_{T}$, $V_{e}$ and B.E. are 
the kinetic energy, central force, LS force, tensor force, Coulomb force and 
total binding energy, respectively. They are in units of MeV. 
$\mbox{\boldmath $L$}^2$ is the expectation value of the orbital angular 
momentum. 
$R_{rms}$ is the root-mean-square radius of the system in units of fm. 
The experimental values of the binding energy and $R_{rms}$ of $^4$He are  
B.E. = 28.3 MeV and $R_{rms}$ = 1.47 fm.\cite{ExpRrms_d_t_4He}}
\begin{center}
\begin{tabular}{l|cccccccc}
\hline \hline
 & $T$ & $V_C$ & $V_{LS}$ & $V_{T}$ & $V_{e}$ & B.E. & 
 $\mbox{\boldmath $L$}^2$ & $R_{rms}$ \\
\hline
Usual AMD     & 56.0 & $-$85.3 & 0.0 & $-$1.7  & 0.9 & 30.1 & 0.01 & 1.37 \\
Isospin fixed & 57.4 & $-$85.7 & 0.2 & $-$4.2  & 0.9 & 31.4 & 0.06 & 1.36 \\
Isospin free  & 77.6 & $-$96.2 & 2.1 & $-$22.4 & 1.0 & 37.9 & 0.41 & 1.22 \\
\hline
\end{tabular}
\end{center}

\end{table}

We consider that these failures are caused by the 
too-large contribution of the central term of the F-T potential. 
Due to the extension of the AMD framework, 
the contribution of the tensor force becomes large, as we expected. 
However, as shown in ``Usual AMD'' in Table II, 
the central force in the F-T potential is so strong that 
$^4$He is already sufficiently bound without any attraction by the 
tensor force. 
As a result of the large contribution of the tensor force gained by 
the extended framework 
in addition to such large contribution of the central force, 
overbinding and shrinkage occur. 
We can explain these failures as being attributed to a mismatching 
between the extended model space of the present framework and the F-T potential. 
The F-T potential should be used 
in a simple model space, as written in the original paper\cite{Furutani}. 
In a simple model space, an explicit tensor term has only a minor effect, 
while the dominant contribution of the original tensor force is 
renormalized into the 
large attraction of the central term in the F-T potential. 
It can well reproduce the total binding energy of $^4$He 
when it is used in a small model space. 
In the case of the present calculation,  
due to the relatively large  model space of the extended AMD, 
the contribution of the tensor force becomes large.  
Consequently, since the tensor force is already renormalized into 
the central force,  
it is doubly counted in the central term and in the direct tensor term, 
causing such overbinding.

\subsection{Different width wave packets and angular-momentum projection}

As explained in the former subsection, 
we can not quantitatively reproduce the properties of the deuteron and $^4$He 
with the F-T potential, which has a strong $^3E$ central force. 
Since this potential is not suitable for our research strategy, where we treat 
the tensor force directly without renormalizing into the central force, 
we employ another type of potential based on  the AK(T) potential, 
in which the magnitude of the $^3E$ central force is nearly equal to 
that of the tensor force. 
We think that the AK(T) potential is more suitable to the extended 
AMD framework than the F-T potential, because the tensor force is 
not renormalized into the central force. 
As discussed in Ref. [\ref{Myo}],
the matrix element of the tensor force 
is reduced approximately by a factor of $\sqrt{\frac{1}{2}}$ in 
simple mean-field approaches.  Therefore, we enhance the strength of  
the tensor force by some factor also in the case of AMD 
in order to properly evaluate the matrix element of the tensor force. 
In the present calculation, we phenomenologically make the strength 
of the tensor force in the original AK(T) potential  
twice enhanced, so as to reproduce the binding energy and radius of $^4$He. 

\begin{table}[b]
\caption{Summary of conditions in calculating $^4$He with the AK(T) potential. 
``$N$'' means the number of wave packets for a single nucleon wave function. 
``$\nu$'s'' indicates that the width parameters of the wave 
packets 
are equal to (``com.'') or different from each other (``diff.''). 
``$\nu_1 \sim \nu_5$'' show values of the adopted width parameters in units 
of fm$^{-2}$. 
``$J$ pro.'' indicates that 
the angular momentum projection has been done (``w'') or not (``w/o''). 
``w/o'' in ``$J$ const.'' means that a solution is obtained 
by the usual frictional cooling method without a $J$ constraint. 
We constructed a better solution by the frictional cooling method 
with the $J$ constraint (``w'' in ``$J$ const.''). 
A detailed explanation is given in the text. 
\label{4He:Akaishi_condition}}
\begin{center}
\begin{tabular}{l|c|cccccc|c|c}
\hline \hline
    & $N$ & $\nu$'s & $\nu_1$ & $\nu_2$ & $\nu_3$ & $\nu_4$ & $\nu_5$ & $J$ 
    pro. & $J$ const. \\
\hline
I   &  4  & com.  & 0.60 & 0.60 & 0.60 & 0.60 & ---  & w/o & w/o\\
II  &  4  & com.  & 0.60 & 0.60 & 0.60 & 0.60 & ---  & w   & w/o\\
III &  4  & diff. & 0.30 & 0.51 & 0.88 & 1.50 & ---  & w/o & w/o\\
IV  &  4  & diff. & 0.30 & 0.51 & 0.88 & 1.50 & ---  & w   & w/o\\
V   &  4  & diff. & 0.30 & 0.51 & 0.88 & 1.50 & ---  & w   & w  \\
V'  &  5  & diff. & 0.30 & 0.45 & 0.67 & 1.00 & 1.50 & w   & w  \\
\hline
\end{tabular}
\end{center}
\end{table}

\begin{table}[b]
\caption{Results of $^4$He with the AK(T) potential calculated under various 
conditions. 
The conditions ``I $\sim$ V'~'' correspond to 
TABLE \ref{4He:Akaishi_condition}. 
$\mbox{\boldmath $J$}^2$, $\mbox{\boldmath $L$}^2$ and 
$\mbox{\boldmath $S$}^2$ 
are the expectation values of the total angular momentum, the orbital angular 
momentum and the spin, respectively. 
\label{4He:Akaishi}}
\begin{center}
\begin{tabular}{l|cccc|cc|ccc}
\hline \hline
 & $T$ & $V_C$ & $V_T$ & $V_e$ & B.E. & $R_{rms}$ & $\mbox{\boldmath $J$}^2$ & 
 $\mbox{\boldmath $L$}^2$ & $\mbox{\boldmath $S$}^2$ \\
\hline
I   & 94.1 & $-$39.4 & $-$66.4 & 0.9 & 10.8 & 1.23 & 0.32 & 1.50 & 1.31 \\
II  & 91.7 & $-$40.3 & $-$68.0 & 0.9 & 15.7 & 1.23 & 0.00 & 1.27 & 1.27 \\
III & 79.0 & $-$41.9 & $-$55.4 & 0.9 & 17.4 & 1.32 & 0.23 & 0.80 & 0.68 \\
IV  & 73.4 & $-$43.5 & $-$56.2 & 0.9 & 25.3 & 1.32 & 0.00 & 0.57 & 0.57 \\
V   & 70.4 & $-$43.5 & $-$56.3 & 0.9 & 28.6 & 1.35 & 0.00 & 0.50 & 0.51 \\
V'  & 69.0 & $-$43.0 & $-$55.5 & 0.9 & 28.7 & 1.35 & 0.00 & 0.54 & 0.54 \\
\hline
\end{tabular}
\end{center}
\end{table}

We calculated $^4$He with the AK(T) potential under various conditions of 
the model space 
and the variation, and  investigated 
the effect of different width wave packets (point iii)) and 
that of the angular momentum projection (an ingredient of the point iv)). 
We list the conditions I $\sim$ V' in Table \ref{4He:Akaishi_condition}. 
Except for case V', the single-nucleon wave functions are represented 
with four wave packets, which corresponds to the case  $N=4$ in 
Eq. (\ref{Eq:SNwfn}). 
We perform the calculation of $^4$He with a common width parameter for 
all packets 
(I, II: $\nu$'s=``com.'') 
and with different width parameters of wave packets 
(III, IV: $\nu$'s=``diff.'').  
In the former case ``com.'', 
a common width parameter $\nu_\alpha=\nu$ is used, and is chosen to be 
$\nu=0.6$ 
so as to minimize the energy of $^4$He. 
In the latter case of different widths (``diff.''), $\nu_\alpha$ is given 
as 
\begin{equation}
\nu_\alpha = \nu_1 \times \left( \nu_N/\nu_1 \right)^{\frac{\alpha-1}{N-1}} .
\end{equation}
After some trials of various sets $\{ \nu_1, \, \nu_4 \}$, 
we found the optimum set, $\nu_1=0.3$ and $\nu_4=1.5$, which 
gives the energy minimum. The adopted width parameters $\{ \nu_\alpha \}$ are 
listed in Table \ref{4He:Akaishi_condition}.  
In each case of a common width (I, II) and different widths (III, IV), 
we first performed the energy variation without the angular-momentum ($J$) 
projection 
(I, III), and projected the obtained wave functions onto the $J$-eigen states 
(II, IV). 

\begin{table}[b]
\caption{Contributions from various $J$ components to the kinetic energy. 
``Before $J$ projection'' corresponds to the III state in 
Table \ref{4He:Akaishi}. 
``overlap'' indicates the ratio of each $(J, J_z)$ component ($| J \, 
J_z \rangle$) 
included in the III state ($| \Phi_{\rm III}\rangle$); 
$|\langle J \, J_z | \Phi_{\rm III} \rangle|^2 / 
\sqrt{\langle J \, J_z |  J \, J_z \rangle \langle \Phi_{\rm III} | 
\Phi_{\rm III} \rangle}$. 
The definitions of $| \Phi_{\rm III}\rangle$ and $| J \, J_z \rangle$ are 
explained 
in the text. 
\label{kinetic}}
\begin{center}
\begin{tabular}{cc|ccc}
\hline \hline 
$J$ & $J_z$ & $T$ & overlap & $T$ $\times$ overlap \\
\hline 
\multicolumn{2}{c|}{Before $J$ projection} &  79.04 & ---   & --- \\
\hline
0     &  0    &  73.43 & 0.955 & 70.09 \\
1     &  1    & 199.75 & 0.007 & 1.38 \\
1     &  0    & ---    & 0.000 & 0.00 \\
1     & $-1$    & 201.24 & 0.007 & 1.35 \\
2     &  2    & 238.19 & 0.005 & 1.23 \\
2     &  1    & 172.96 & 0.002 & 0.31 \\
2     &  0    & ---    & 0.000 & 0.00 \\
2     & $-1$    & 177.75 & 0.002 & 0.30 \\
2     & $-2$    & 238.38 & 0.005 & 1.24 \\
\hline
\end{tabular}
\end{center}
\end{table}

The results of $^4$He calculated under each condition are given in 
Table \ref{4He:Akaishi}. 
By comparing II (a common width) and IV (different widths), it is found 
that 
the reproduction of both the binding energy and the radius is 
improved in result IV.   
This shows the importance of using wave packets with different widths. 

The comparison of III (without $J$ projection) and IV (with $J$ projection) 
indicates 
the importance of the angular-momentum projection. 
As can be seen in Table \ref{4He:Akaishi}, the binding energy gains 8 MeV in  
result IV 
by the $J$ projection. 
Before the $J$ projection,   
the $\mbox{\boldmath $J$}^2$ value is small, as can be seen in result III, 
which indicates 
that the $J \neq 0$ components are very small.  
Considering that the state is almost the $J=0$ eigen state already before 
the $J$ projection, 
the energy gain by the projection seems to be unexpectedly large. 
This gain is dominantly attributed to a decrease of the kinetic energy term. 
Dissolving the III state ($| \Phi_{\rm III} \rangle$)  
into various $(J, J_z)$ components ($| J \, J_z \rangle$), we analyzed 
the contributions of each $(J, J_z)$ component to the kinetic energy 
(Table \ref{kinetic}). Detailed expressions of $| \Phi_{\rm III} \rangle$ 
and $| J \, J_z \rangle$ are 
\begin{eqnarray}
|\Phi_{\rm III} \rangle \; & \equiv & \; P_{Tz=0} \, P_{{P}=+} \, |
\Phi \rangle, \\
| J \, J_z \rangle \; & \equiv & \; P_{J,Jz} \, |\Phi_{\rm III} \rangle, 
\end{eqnarray}
where $|\Phi \rangle$ is the intrinsic state of the III state. 
As shown in Table \ref{kinetic}, the $J=0$ component is dominant as 
96\%, while the $J=1$ and $J=2$ components are 
mixed into the III state at the rate of a few percent. 
Although the $J=1$ and 2 components are very small, 
they have huge kinetic energy ($\simeq 200$ MeV), and hence the significant 
gain of 
the kinetic energy is caused by excluding the $J\neq0$ components after the 
$J$ projection.
Such a huge kinetic energy of the $J\neq0$ components 
arise from very narrow wave packets, such as  $\nu_4=1.5$ used 
in calculating III and IV. 
It is considered that high-momentum components are contained 
in the $J\neq0$ components of the III state, due to such narrow wave packets. 

Since our calculation procedure is ``Variation Before Projection'' 
with respect to the $J$ projection, 
we can expect to find a better solution of the $J=0$ state than IV. 
For this aim, we apply a constraint condition method in the present framework. 
We put a constraint on the squared total angular momentum 
by setting $W_2=\langle \hat{\mbox{\boldmath $J$}}^2 \rangle$ in the 
Eq. (\ref{constraint}) 
for the energy variation, 
and constructed the states with various values of 
$\langle \hat{\mbox{\boldmath $J$}}^2 \rangle$ ($\hat{\mbox{\boldmath $J$}}$ 
is the total 
angular momentum operator). 
After the energy variation, we projected them onto the $J=0$ state 
and obtained an energy surface as a function of the value of the constraint 
$\langle \hat{\mbox{\boldmath $J$}}^2 \rangle$. 
We then chose the energetically best solution, which is state ``V'' 
shown in Table \ref{4He:Akaishi}. 
In this solution, the binding energy and the radius 
are more improved than IV. 
In this final result, V,
the binding energy and the radius are $-28.6$ MeV and 1.35 fm, respectively. 
We checked the convergence of the solution with an increase of the number 
of wave packets, $N$. 
V' is the result obtained by using five wave packets for each 
nucleon wave function, namely $N=5$ in Eq. (\ref{Eq:SNwfn}), 
in the same way as the case of V for $N=4$. 
Since the changes of various properties from the V state to the V' state 
are very small, 
we consider that the convergence of the solution is sufficient at $N=4$. 

Thus, we confirmed that 
extension iii) {\it different width wave packets} 
and especially the {\it angular-momentum projection} in iv) 
are effective as well as  extensions i) and ii).

\subsection{Systematics of deuteron, triton and alpha with AK(T) potential}

Within the extended AMD framework   
we could rather quantitatively reproduce  the properties of $^4$He 
by using  the AK(T) potential with a twice-enhanced strength of the tensor term. 
We now consider our calculation of the deuteron and triton in the same way 
as in case V for $^4$He. 
The results are given in Table \ref{d_t_4He:Akaishi}.  
Comparing the theoretical values of binding energies and radii with the 
experimental data, we can tolerate the result of the triton, though it is 
somewhat overbound, whereas the deuteron is terribly overbound and overshrunk. 

\begin{table}[b]
\caption{Results of the deuteron, triton and $^4$He 
calculated with AK(T) potential. 
``$N$'' means the number of wave packets for a single nucleon wave function. 
``$x_T$'' indicates the enhancement factor of the strength of the tensor 
force. 
``a$_1$'', ``b'' and ``c'' show the results of the deuteron, triton 
and $^4$He calculated with the twice-enhanced tensor force. 
``a$_2$'' is the deuteron calculated with a 1.5-times enhanced tensor force.  
The experimental values of the binding energy and radius: 
B.E. = 2.2 MeV and $R_{rms} = 1.94$ fm for deuteron, and 
8.5 MeV and $R_{rms} = 1.55$ fm for triton.\cite{ExpRrms_d_t_4He} 
\label{d_t_4He:Akaishi}}
\begin{center}
\begin{tabular}{lc|c|c|ccccccccc}
\hline \hline
 & &  $N$ & $x_T$ &$T$ & $V_C$ & $V_T$ & $V_e$ & B.E. & $R_{rms}$ & 
 $\mbox{\boldmath $J$}^2$ & $\mbox{\boldmath $L$}^2$ & 
 $\mbox{\boldmath $S$}^2$ \\
\hline
a$_1$ & $^2$H & 5 & 2.0  & 34.9 & $-$4.1  & $-$44.1 & 0.0 & 13.4 & 1.00 & 
2.00 & 0.72 & 1.98 \\
a$_2$ & $^2$H & 5 & 1.5  & 23.2 & $-$4.2  & $-$20.7 & 0.0 & 1.7  & 1.19 & 
2.00 & 0.49 & 1.99 \\
b     & $^3$H & 5 & 2.0  & 48.0 & $-$20.5 & $-$38.0 & 0.0 & 10.5 & 1.27 & 
0.75 & 0.54 & 1.01 \\
c     & $^4$He & 4 & 2.0  & 70.4 & $-$43.5 & $-$56.3 & 0.9 & 28.6 & 1.35 & 
0.00 & 0.50 & 0.51 \\
\hline
\end{tabular}
\end{center}
\end{table}

One of the reasons for this failure in systematic reproduction 
is conjectured to be as follows: 
The deuteron is a special system, because it is composed of 
only two nucleons. 
It has just one degree of freedom in the spatial coordinate, namely, 
the relative motion between them. 
Since its relative motion may be described sufficiently well 
by the superposition of Gaussian wave packets in the extended AMD, 
the two-body correlation in the deuteron is incorporated much better than that 
in $^4$He. 
As mentioned in the previous section, Ref. [\ref{Myo}] pointed out that 
the matrix element of the tensor force is reduced approximately by 
a factor of  $\sqrt{\frac{1}{2}}$ 
in the simple mean-field type treatment of $^4$He due to the limitation 
of the model space, which is insufficient to incorporate a two-body 
correlation. 
In order to regain this reduction, we need a twice-stronger tensor force 
for the 
calculation of $^4$He. However, because of the better description of 
the two-body correlation 
in the deuteron wave function than  in $^4$He, the reduction factor of 
the matrix 
element must be smaller and the enhancement factor of the strength should be 
smaller in the deuteron than  in $^4$He. 
This means that 
the twice-enhanced tensor force is too strong for the deuteron. 
Actually, from the value $\mbox{\boldmath $L$}^2=0.72$ shown in  
row ``a$_1$'',  
we can find that the $D$-state probability of the deuteron is too large 
(about 12\%),  
compared with the well-known value of several \%. 
We then calculated the deuteron with a smaller 
enhancement factor, $x_T=1.5$, of 
the strength of the tensor force. The result is shown in row 
``a$_2$'' of Table \ref{d_t_4He:Akaishi}. 
In this case, $\mbox{\boldmath $L$}^2=0.49$,  
which indicates that the $D$-state probability is estimated to be 
a reasonable value as 8\%. 
The binding energy also decreases to be 1.7 MeV, which well agrees 
with the experimental data.  
These results support that an about 
1.5 time enhancement of the tensor force is suitable to 
reproduce structure properties in the case of the deuteron system.

\section{Analysis of and discussion on single particle levels}

The $^4$He obtained by the extended version of AMD with the AK(T) potential 
has a significant contribution of the tensor force
to the potential energy. 
Due to the tensor force, the obtained wave function is expected to be more 
complicated than the ordinary $(0s)^4$-like wave function 
obtained with the usual effective forces with no tensor term. 
In this section, we clarify its internal structure. 

\begin{table}[b]
\caption{Single particle levels of $^4$He in the state V. 
``S.P.E.'' is the single particle energy of each level in unit of MeV. 
``$\mbox{\boldmath $j$}^2$'' and ``$\mbox{\boldmath $l$}^2$'' are the 
expectation value of total angular momentum and orbital angular momentum, 
respectively. ``P(+) (P($-$))'' and ``P(proton) (P(neutron))'' are the 
ratio of 
the positive(negative) parity component and the proton(neutron) component 
included in each level, 
respectively. They are in unit of \%. 
\label{SPlevels}}
\begin{center}
\begin{tabular}{lccccccc}
\hline \hline
Level & S.P.E. & $\mbox{\boldmath $j$}^2$ & $\mbox{\boldmath $l$}^2$ & P(+) 
& P($-$) 
& P(proton) & P(neutron)\\
\hline
1   & $-$4.04 & 1.16 & 0.83 & 81 & 19 & 61 & 39 \\
2   & $-$4.52 & 1.17 & 0.85 & 81 & 19 & 59 & 41 \\
3   & $-$8.28 & 1.09 & 0.68 & 85 & 15 & 30 & 70 \\
4   & $-$8.78 & 1.06 & 0.65 & 86 & 14 & 45 & 55 \\
\hline
\end{tabular}
\end{center}
\end{table}

In the procedure of the present calculation to obtain a total wave function, 
we operate parity-, charge- and angular-momentum-projections   
to an intrinsic wave function.  
The wave function of $^4$He is written as   
\begin{eqnarray}
|^4{\rm He} \rangle \; = \; P_{J=0} \, P_{T_z=0} \, P_{{\cal P}=+} \; | 
\Phi \rangle,    
\end{eqnarray}
where 
the intrinsic state, $| \Phi \rangle$, is expressed by a single Slater 
determinant. 
We can analyze the Hartree-Fock-like single-particle orbits, $| \alpha_i 
\rangle$ 
($i=1 \sim 4$), (AMD-HF orbits) 
extracted from the AMD wave function, $| \Phi \rangle$, 
by the method given in Ref. [\ref{AMD-HF1}]. 

Table \ref{SPlevels} shows the properties of the AMD-HF levels extracted from 
the intrinsic wave function of result V in Table \ref{4He:Akaishi}. 
Apparently, the intrinsic state $|\Phi \rangle$ 
is not the simple $(0s)^4$ state. If $|\Phi \rangle$ were the $(0s)^4$ 
state, 
$\mbox{\boldmath $j$}^2$, $\mbox{\boldmath $l$}^2$ and $P(+)$ should be 
equal to 
0.25, 0.00 and 100\% at each level, respectively. The obtained values, 
however, 
deviate from these values. 
The obtained levels contain a negative-parity component with $15 
\sim 20 \%$, 
and $\mbox{\boldmath $l$}^2$ is non-zero. 
These results indicate a mixture of the $p$ state in all single-particle 
orbits. It is interesting that 
they are also proton-neutron-mixed states, as shown in Table \ref{SPlevels}.

To investigate in more detail, we have decomposed the AMD-HF level 
into the eigen-state of parity and charge. 
The composition of each level is shown in Table \ref{SPcomposition}. 
$\mbox{\boldmath $l$}^2$ and 
the root-mean-square radius of each component are given in Table 
\ref{SPcompL2} and \ref{SPcompR}, respectively. 
The negative-parity components included in all AMD-HF levels are found to 
be dominated by $p$ orbits, because $\mbox{\boldmath $l$}^2$ for the 
negative-parity component 
for both of proton and neutron is nearly equal to 2, as shown in 
Table \ref{SPcompL2}.  
In the positive-parity components, non-zero $\mbox{\boldmath $l$}^2$ 
values imply the mixing of $d$ orbits with the dominant $s$ orbit. 
As shown in Table \ref{SPcompR}, the radius of the negative-parity components 
is about 1.28 fm. If we assume a harmonic-oscillator 
wave function, this radius, 1.28 fm, of the $p$-orbits 
suggests an unusually small oscillator size. 
On the other hand, the radius of the positive-parity components is 
about 1.6 fm, 
which suggests a normal size for the $s$ obit in $^4$He.  
Thus, it is considered that 
the {\it narrow p state} (negative-parity component) is mixed 
with the normal-size $s$ state (positive-parity component) in each level.

\begin{table}[b]
\caption{Composition of the single-particle levels. 
proton (neutron) + ($-$) means the component of the 
proton (neutron) with the positive (negative) parity. 
The values are in unit of \%. 
\label{SPcomposition}}
\begin{center}
\begin{tabular}{lcccc}
\hline \hline
Level & proton + & neutron + & proton $-$  & neutron $-$  \\
\hline
1   & 55 & 26 & 6 & 13 \\
2   & 50 & 30 & 9 & 10 \\
3   & 21 & 64 & 9 & 6  \\
4   & 37 & 49 & 8 & 6  \\
\hline
\end{tabular}
\end{center}
\end{table}

\begin{table}[b]
\caption{Expectation value of $\mbox{\boldmath $l$}^2$ 
of each component. 
\label{SPcompL2}}
\begin{center}
\begin{tabular}{lcccc}
\hline \hline
Level & proton + & neutron + & proton $-$  & neutron $-$  \\
\hline
1   & 0.16 & 0.99 & 2.58 & 2.56 \\
2   & 0.29 & 0.64 & 2.61 & 2.61 \\
3   & 1.09 & 0.10 & 2.68 & 2.63  \\
4   & 0.41 & 0.26 & 2.67 & 2.48  \\
\hline
\end{tabular}
\end{center}
\end{table}

\begin{table}[b]
\caption{Root-mean-square radius of each component. 
The units are fm. 
\label{SPcompR}}
\begin{center}
\begin{tabular}{lcccc}
\hline \hline
Level & proton + & neutron + & proton $-$  & neutron $-$  \\
\hline
1   & 1.59 & 1.70 & 1.30 & 1.27 \\
2   & 1.58 & 1.60 & 1.27 & 1.31 \\
3   & 1.55 & 1.54 & 1.26 & 1.28  \\
4   & 1.52 & 1.51 & 1.28 & 1.25  \\
\hline
\end{tabular}
\end{center}
\end{table}

\begin{figure}[t]
\caption{Overlaps between the AMD-HF states extracted from state V in 
Table \ref{4He:Akaishi} 
and the $s$ and $p$ states in the harmonic-oscillator potential 
with various width parameters ($\nu_s$ and $\nu_p$ fm$^{-2}$). 
(a) Overlap between the 
positive parity component of each AMD-HF state and the $s$ state; 
$|\langle {\rm s_{H.O.}}| \alpha_i^+ \rangle|^2$.   
(b) Overlap between the 
negative parity component of each AMD-HF state and the $p$ state; 
$|\langle {\rm p_{H.O.}}| \alpha_i^- \rangle|^2$. 
The four lines in each figure correspond to four single-particle levels, 
respectively.   
 \label{SPov_4He_Ak}}
\begin{center}
\begin{minipage}[t]{6.5cm}
\begin{center}
\includegraphics[width=6.5cm]{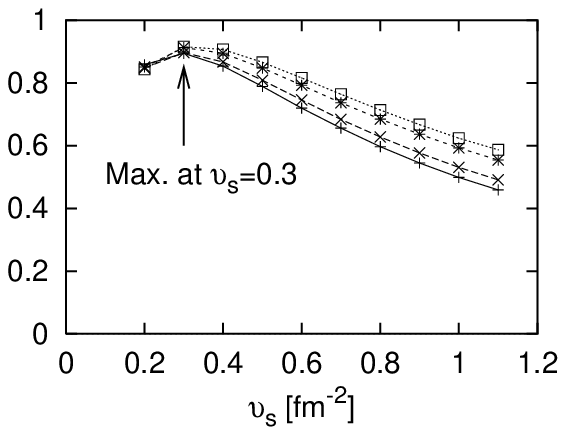}\\
(a) positive parity
\end{center}
\end{minipage}
\hspace{0.5cm}
\begin{minipage}[t]{6.5cm}
\begin{center}
\includegraphics[width=6.5cm]{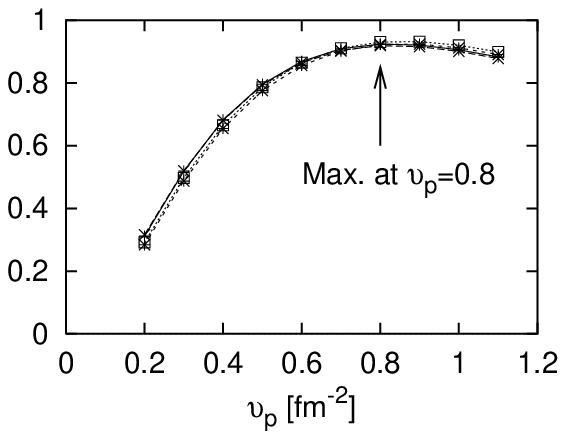}\\
(b) negative parity
\end{center}
\end{minipage}
\end{center}
\end{figure}

\begin{table}[b]
\caption{Composition of the $p$ state at $\nu_p=0.8$ fm$^{-2}$. 
The units are \%. 
``p'' and ``n'' in the column ``p or n'' mean the proton and neutron 
component, respectively. ``p+n'' means the sum of proton and neutron 
components.
\label{AnaHOpot}}
\begin{center}
\begin{tabular}{lc|cccc}
\hline \hline
 & p or n & level 1 & level 2  & level 3 & level 4  \\
\hline
$0p_\frac{1}{2}$ & p  & 27.2 & 41.3 & 54.8 & 48.0 \\
$0p_\frac{1}{2}$ & n  & 59.8 & 43.6 & 32.4 & 39.2 \\
$0p_\frac{1}{2}$ & p+n  & 87.0 & 84.9 & 87.2 & 87.3 \\
\hline
$0p_\frac{3}{2}$ & p  & 2.0 & 2.9 & 3.3 & 3.6 \\
$0p_\frac{3}{2}$ & n  & 3.4 & 4.0 & 1.8 & 2.2 \\
$0p_\frac{3}{2}$ & p+n  & 5.4 & 6.9 & 5.1 & 5.8 \\
\hline
$0p$ & p+n  & 92.4 & 91.8 & 92.3 & 93.1 \\
\hline
\end{tabular}
\end{center}
\end{table}

In order to confirm  the above picture, 
we analyze the AMD-HF orbits with 
simple $s$- and $p$-orbits written in terms of 
harmonic-oscillator (H.O.) wave functions:  
\begin{eqnarray}
|s_{\rm H.O.} \rangle : \hspace{0.5cm} & \langle \mbox{\boldmath $r$} | 
0s_\frac{1}{2}, m, \tau_z\rangle &  =  
 R_s (r) \left| \frac{1}{2}, m \right\rangle 
\cdot \left| \frac{1}{2}, \tau_z \right\rangle, \\
& & R_s (r) = N_s^{-\frac{1}{2}}  \exp \left[ -\nu_s r^2 \right], 
\label{HO_s}\\
|p_{\rm H.O.} \rangle : \hspace{0.5cm} & \langle \mbox{\boldmath $r$} | 
0p_j, m, \tau_z\rangle & =  
R_p (r) \left[ \sum_{m_l, m_s}  \left( 1 m_l, \frac{1}{2} m_s | j m \right) 
Y_{1 m_l} (\Omega) 
\left| \frac{1}{2}, m_s \right\rangle \right] \cdot \left| \frac{1}{2}, 
\tau_z \right\rangle, \\
& & R_p (r) = N_p^{-\frac{1}{2}} \, r \, \exp \left[ -\nu_p r^2 \right] . 
\label{HO_p}
\end{eqnarray}
Here, $\nu_s$ and $\nu_p$ indicate the width parameters of the $s$ and $p$ 
orbits, respectively. 
Fig. \ref{SPov_4He_Ak}(a) (\ref{SPov_4He_Ak}(b)) 
shows the overlap between the positive (negative)-parity component of 
the AMD-HF states, 
$| \alpha_i^\pm \rangle$,  
and the $s_{\rm H.O.}$ ($p_{\rm H.O.}$) state as a function of the H.O. 
width parameter, $\nu_s$ ($\nu_p$). Here, $| \alpha_i^\pm \rangle$ is 
the parity-eigen state projected from $| \alpha_i \rangle$, where 
$\langle \alpha_i^\pm | \alpha_i^\pm \rangle=1$.   
$|s_{\rm H.O.}\rangle$ and $|p_{\rm H.O.} \rangle$ have maximum overlap 
(92\%) with 
$| \alpha_i^+ \rangle$ and $| \alpha_i^- \rangle$ at $\nu_s=0.3$ fm$^{-2}$ and 
$\nu_p=0.8$ fm$^{-2}$, respectively. 
On the other hand, the typical width parameter, $\nu=0.26$ fm$^{-2}$, is known 
to reproduce the size of $^4$He by the H.O. $(0s)^4$ wave function.  
Thus, it is explicitly confirmed that 
the narrow $p$ state is mixed with the normal-size $s$ state. 
It is worth showing that 
the 92\% $0p$ state consists of  86\% $0p_\frac{1}{2}$ and 
6\% $0p_\frac{3}{2}$ (Table \ref{AnaHOpot}).

\begin{figure}[t]
\caption{Comparison of the radial dependence of the single-particle 
wave functions and the tensor force. ``$R_s * R_p$'' (dashed line) is 
the product of the radial wave function of the $s$ state (Eq. \ref{HO_s}) 
and that of the $p$ state (Eq. \ref{HO_p}). ``$V_T$'' (solid line) 
is the radial 
part of the tensor force of the AK(T) potential. 
Both are depicted after being multiplied by $r^2$. \label{HOsp_vs_VtAK}}
\begin{center}
\includegraphics[width=7cm]{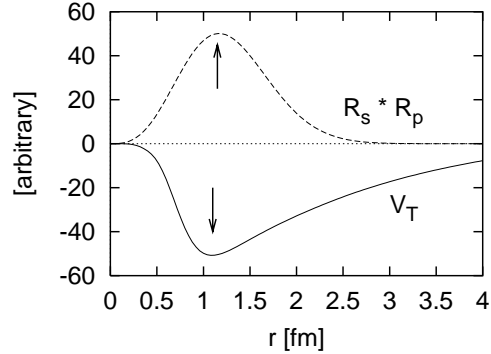}
\end{center}
\end{figure}
 
The reason for the mixture of such a shrunk $p$ orbit can be 
understood as follows. 
In the single-particle picture, the matrix element of the tensor force 
is gained by the coupling of the $s$ state and the $p$ state, such as 
$\langle \, [pp]d \, | v^T_{12} | \, [ss]s \, \rangle$ in which 
two nucleons in the $p$ orbit and those in the $s$ orbit are coupled to 
the $d$ orbit and the $s$ orbit, respectively. 
Fig. \ref{HOsp_vs_VtAK} shows the product of the radial wave functions of 
the $s_{\rm H.O.}$ orbit ($\nu_s=0.3$) and the narrow $p_{\rm H.O.}$ orbit 
($\nu_p=0.8$), 
and also the radial part of the tensor force in the AK(T) potential. 
The peak position of the product of these two orbits is consistent with 
the potential range of the tensor force.  
Such a combination as $s_{\rm H.O.}(\nu_s=0.3)$ and $p_{\rm H.O.}(\nu_p=0.8)$ 
is suitable to incorporate the tensor effect into the binding energy. 
As a result of the analysis of the single particle levels, 
we are convinced of the role of extension iii). In other words,  
we should use wave packets with different width parameters in order to 
describe the shrunk $p$ orbit simultaneously with the normal-size $s$ orbit.

\section{Summary}

In the present paper, we tried to construct a new model 
that can treat the tensor force directly, based on the antisymmetrized 
molecular dynamics method (AMD). 
We employed the Furutani-Tamagaki potential (F-T potential) and 
the Akaishi(T) potential 
(AK(T) potential).  
As a result of practical calculations, we confirmed that 
four extensions in the AMD framework are effective for a direct treatment of 
the tensor force: 
i) a parity-violating mean field and a strong correlation between spin and space,  
ii) flexibility on the isospin wave function, 
iii) different width wave packets
and  
iv) charge-, parity- and angular-momentum-projections. 
Due to all of the above extensions, we obtained a satisfactory solution of 
$^4$He.  
We can reproduce the binding energy for $^4$He 
with the AK(T) potential with a twice-enhanced tensor strength; also,  
the radius of the calculated $^4$He well agrees with the experimental value; 
The binding energy and root-mean-square radius were calculated to be 
28.6 MeV and 1.35 fm, respectively. 
The energy composition is as follows: 
$T=70.4$ MeV, $V_C=-43.5$ MeV, $V_T=-56.3$ MeV and  $V_{Coulomb}=0.9$ MeV.  
Thus, this solution has a large contribution of the tensor force, as we 
expected.     
We found that the influence of the large contribution of the tensor force 
appeared in single-particle wave functions of this $^4$He solution.  
The shrunk $p$ orbit was mixed to the usual $0s$ orbit to gain the tensor force. 

We could gain a large contribution of tensor force 
in the deuteron, triton and $^4$He with the extended version of AMD.
However, we are not satisfied with the results 
because we need a two-fold enhancement of the tensor force in the AK(T) potential 
to quantitatively describe the properties of $^4$He, and have not yet
succeeded to reproduce the systematics of the binding energies of 
the deuteron, triton, and $^4$He.
These fact may mean that the extension of the framework is still insufficient. 
Referring to the deuteron system,
we here comment on possible residual correlations.
Since the deuteron has only one degree of freedom, i.e. the relative motion, 
the description of the two-body correlation  
is considered to be directly improved by increasing the number of 
wave-packets.
In fact, we have found that the tensor contribution in the deuteron 
changes drastically with an increase of the number of wave packets 
from $N=5$ to 10, with its binding energy changing from 13.4 MeV to 18.2 MeV.
It is different from
the case of $^4$He, where the results almost converge at $N=4$.
In the case of $N=10$ for the deuteron, only a 1.4-times enhancement of the 
tensor force is sufficient to get the satisfactory result of 1.6 MeV for  
the deuteron binding energy, which  
agrees with the experimental data (2.2 MeV).
This result implies that there is some scope to improve 
the description of the two-body correlation in our framework. 
In a mean-field-type model, such as the AMD method, 
it is difficult to treat the short-range part of the tensor force. 
This difficulty is expected to be overcome by introducing 
the so-called $T$-type base,\cite{VT-Hyblid:Ikeda} 
or renormalizing such a part of the tensor force into the central force. 
In order to perform further studies based on AMD, the latter way is suitable 
for our method. 
Due to renormalization of the short-range tensor force, the central force 
is expected to become more attractive than the present one. 
From the result obtained with the F-T potential, which has a strong central force, 
we can conjecture that the binding energy is easily reproduced 
by the renormalized force without the enhancement factor of the tensor force.  
In a study with a mean-field approach by Sugimoto {\it et al.}
\cite{Sugimoto:CPPHF}, they use 
the enhancement factor 1.5 for the same tensor force as in the present study.
By adjusting the strength of the central force, 
the binding energy of $^{4}$He is well described in their study.
Although the central force is phenomenologically fitted,
their approach is interpreted to simulate the above-mentioned 
renormalization of the short-range tensor force in the central force.
Needless to say, it is important to derive effective nuclear interactions 
with a direct long-range part and a renormalized short-range part of the
tensor force based on realistic forces, as much as possible.
Such a renormalization framework may work effectively to   
investigate the influence of the tensor force on the nuclear structure,  
because the short-range part of tensor force is expected to affect 
mainly the bulk properties of nuclei, while its long-range part 
should be sensitive to the 
details of the nuclear structure.

\section*{Acknowledgements}

One of the authors (A. D.) thanks Dr. S. Sugimoto for fruitful discussions 
and his encouragement.  
He is very thankful to Prof. A. Tohsaki and Dr. M. Kimura 
for their advice on a mathematical technique. 
This work is supported by JSPS Research Fellowships for Young Scientists,  
and was performed under the Research Project for Study of 
Unstable Nuclei from Nuclear Cluster Aspect, sponsored by the 
Institute of Physical and Chemical Research (RIKEN). 

\appendix
\section{} 

Parameters in the Akaishi(T) potential are given in Table \ref{data_Akaishi_pot}. 
$C^{1E}_n$, $C^{3E}_n$,  $C^{1O}_n$  and  $C^{3O}_n$ are 
parameters of the central force in Eq. (\ref{Vc}) and 
$T^{3E}_n$ is that of the tensor force in Eq. (\ref{Vt}).
Each range parameter, $b_n$, is generated from $b_1$ 
and $b_{10}$ by $b_n=b_1 \times (b_{10}/b_1)^{(n-1)/9}$.

\begin{table}[h]
\caption{Range parameters $\{b_n\}$ and coefficients $\{C^{X}_n\}$ in 
$v^C_{ij}$ 
and $\{T^{X}_n\}$ in $v^T_{ij}$ of AK(T) potential. 
The units of $\{b_n\}$, $\{C^{X}_n\}$ and $\{T^{X}_n\}$ are fm, MeV and 
MeV/fm$^2$, 
respectively. \label{data_Akaishi_pot}
The definitions are given in Eqs. (\ref{Vc}) and (\ref{Vt}).}
\begin{tabular}[t]{c|c|cccc|c}
\hline \hline
$n$ & $b_n$ & $C^{1E}_n$ & $C^{3E}_n$ & $C^{1O}_n$ & $C^{3O}_n$ & $T^{3E}_n$ 
\\
\hline
1  & 0.1600 & $-$3.4970$\times10^2$ & $-$3.2879$\times10^2$ &  
2.7042$\times10^1$ &  1.5355$\times10^2$ & $-$2.2053$\times10^2$ \\
2  & 0.2141 &  1.4440$\times10^3$ &  1.2518$\times10^3$ & 
$-$2.0400$\times10^2$ & $-$1.2405$\times10^2$ &  1.1121$\times10^3$ \\
3  & 0.2865 & $-$2.9647$\times10^3$ & $-$2.5066$\times10^3$ &  
9.0073$\times10^2$ & $-$5.1523$\times10^2$ & $-$1.6079$\times10^3$ \\
4  & 0.3833 &  1.4499$\times10^3$ &  1.3850$\times10^3$ & 
$-$1.8442$\times10^3$ &  4.7590$\times10^2$ &  1.3446$\times10^2$ \\
5  & 0.5129 &  1.6601$\times10^3$ &  1.2147$\times10^3$ &  
1.3668$\times10^3$ &  4.3360$\times10^2$ & $-$3.8447$\times10^2$ \\
6  & 0.6863 & $-$7.7942$\times10^2$ & $-$6.5449$\times10^2$ &  
1.8709$\times10^1$ &  2.5416$\times10^1$ & $-$1.8164$\times10^2$ \\
7  & 0.9183 & $-$1.4783$\times10^1$ &  7.6192$\times10^1$ &  
8.5685$\times10^1$ & $-$4.6085$\times10^1$ & $-$4.1188$\times10^1$ \\
8  & 1.229 & $-$6.1748$\times10^1$ & $-$6.5063$\times10^1$ & 
$-$3.5048$\times10^{-1}$ &  4.3151 & $-$1.2277$\times10^1$ \\
9  & 1.644 &  6.5369 &  1.1420$\times10^1$ & $-$6.3206 & $-$6.8892 & 
$-$2.2104 \\
10 & 2.200 & $-$6.4642 & $-$6.6031 &  1.3007$\times10^1$ &  1.6804 & 
$-$6.5291$\times10^{-1}$ \\
\hline
\end{tabular}
\end{table}

%

\end{document}